\documentclass[aps,twocolumn,superscriptaddress,prb,showpacs]{revtex4}
\usepackage{graphicx}
\pdfoutput=1
\newcommand{\AP}{Applied Physics, Chalmers University of Technology, SE-412 96 G{\"o}teborg, Sweden}
\newcommand{\MCtwo}{Microtechnology and Nanoscience, MC2, 
Chalmers University of Technology, SE-412 96 G{\"o}teborg, Sweden}

\newcommand{\vdW}{{\mbox{\scriptsize vdW-DF}}}
\newcommand{\GGA}{{\mbox{\scriptsize GGA}}}
\newcommand{\LDA}{{\mbox{\scriptsize LDA}}}
\newcommand{\PBE}{{\mbox{\scriptsize PBE}}}
\newcommand{\tot}{{\mbox{\scriptsize tot}}}
\newcommand{\nl}{{\mbox{\scriptsize nl}}}
\newcommand{\floor}{{\mbox{\scriptsize floor}}}
\newcommand{\reff}{{\mbox{\scriptsize ref}}}
\newcommand{\dac}{{\mbox{\scriptsize dac}}}

\begin{document}

\title{Role of van der Waals bonding in layered oxide: Bulk vanadium pentoxide}

\author{Elisa Londero}\affiliation{\MCtwo}\affiliation{\AP}
\author{Elsebeth Schr{\"o}der}\thanks{Corresponding author}\email{schroder@chalmers.se}%
\affiliation{\MCtwo}

\date{June 5, 2010}

%%%%%%%%%%%%%%%%%%%%%%%%%%%%%%%%%%%%%%%%%%%%%%%%%%%%%%%%%%%%%%%%%%%%%%%%%%%%%%%%%%%%%%%%%%%%%%
\begin{abstract} 
Sparse matter is characterized by regions with low electron density and its understanding 
calls for methods to accurately calculate both the van der Waals (vdW) interactions and other 
bonding. Here we present a first-principles density functional theory (DFT) study of a layered oxide (V$_2$O$_5$) 
bulk structure which shows charge voids in between the layers and we highlight the role of the 
vdW forces in building up material cohesion. 
The result of previous first-principles studies involving semilocal approximations to 
the exchange-correlation functional in DFT gave results in 
good agreement with experiments for the two in-plane lattice parameters of the unit cell but 
overestimated the parameter for the stacking direction. To recover the third parameter we include the 
nonlocal (dispersive) vdW interactions through the vdW-DF method [Dion et al.,
Phys.\ Rev.\ Lett.\ \textbf{92}, 246401 (2004)] testing also various choices of exchange flavors. 
We find that the transferable 
first-principle vdW-DF calculations stabilizes the bulk structure. The vdW-DF method gives results 
in fairly good agreement with experiments for all three lattice parameters.
\end{abstract}

\maketitle

%%%%%%%%%%%%%%%%%%%%%%%%%%%%%%%%%%%%%%%%%%%%%%%%%%%%%%%%%%%%%%%%%%%%%%%%%%%%%%%%%%%%%%%%%%%%%%
\section{Introduction} 
Vanadium is one of the most abundant metals on earth and it is found in about 
150 different minerals. 
Catalysts based on vanadium oxides are widely used in the production of chemicals and 
in the reduction of environmental pollution, in particular the vanadium pentoxide (V$_2$O$_5$) form is 
extensively used.\cite{Keller} 
In recent years V$_2$O$_5$ has also been used with intercalating Li ions
for high-capacity solid-state batteries.\cite{Julien,Sabana,Chou,Braithwaite} 
The applied-research and industrial focus on catalysis 
and batteries involving V$_2$O$_5$ has led to a substantial amount of atomic-scale theory studies focusing on the 
V$_2$O$_5$ bulk\cite{Braithwaite,Pirovano,Pirovano2007,willinger2004,GPMV,Goclon,Xiao,Reeswinkel}
and its surfaces.\cite{Pirovano,Kresse01,Li,Goclon2,Hejduk}  

The bulk of V$_2$O$_5$ has a stacked structure. Previous atomic-scale calculations that were based on 
density functional theory (DFT) with the popular semilocal 
generalized gradient approximations (GGA) have often failed in arriving at   
the experimentally known value of the lattice parameter
in the stacking direction of the layers.\cite{Pirovano,Pirovano2007,willinger2004,Goclon,Xiao,Kresse01} 
This is believed to happen because GGA does not describe the van der Waals (vdW) forces,\cite{layerPRL} 
and because these are expected to play an 
important role in binding the layers in V$_2$O$_5$ bulk.\cite{Pirovano} 
In several GGA-based V$_2$O$_5$ surface or vacancy studies this
deficiency of GGA was either ignored or worked around by imposing the experimentally obtained
lattice parameter or unit cell volume. Other groups have employed the semi-empirical method
DFT-D (Ref.\ \onlinecite{grimme}) for adding the vdW forces.\cite{sauer2008}

In this paper we use the first-principles vdW density-functional approach, vdW-DF,\cite{Dion,Thonhauser} to  
determine the V$_2$O$_5$ bulk structure. We find that the vdW forces are indeed  
important for binding the structure, and we analyse the binding
within and mainly between the layers. We here focus solely on the common $\alpha$-V$_2$O$_5$
bulk structure and ignore the more exotic $\gamma$-V$_2$O$_5$ structure.\cite{willinger2004,GPMV} 

The vdW-DF method has a different philosophy and aim and some advantages over the DFT-D method.
The DFT-D approach to include the vdW forces is an atom-centered
pair-potentials method. 
The vdW-DF method is directly based on the electron response. 
The vdW-DF method correctly describes the interaction as arising in the tails of the 
electron distribution, not at the atomic centers.
Unlike DFT-D, vdW-DF provides a framework which is well suited to include effects of
image planes.\cite{NTgg}
The DFT-D omission of image planes can result in inconsistencies of the description
across a range of distances.\cite{berland2010}
This effect could be important in materials like V$_2$O$_5$ where the corrugation
is large, and where as a result both relatively small and larger distances contribute
to the vdW interaction simultaneously.  

The outline is as follows. In Sec.\ II we describe the structure of V$_2$O$_5$
as known from experiments. 
In Sec.\ III we describe the computational methods used, both the self-consistent (sc)
GGA calculations that provide the electron density, and the postprocess procedure that 
takes this density as input for obtaining the vdW contribution. 
Sec.\ IV is devoted to the discussion of our results and Sec.\ V includes a summary. 
%%%%%%%%%%%%%%%%%%%%%%%%%%%%%%%%%%%%%%%%%%%%%%%%%%%%%%%%%%%%%%%%%%%%%%%%%%%%%%%%%%%%%%%%%%%%%%

\section{Material structure}
\begin{figure}[bt]
\begin{center}
\includegraphics[width=0.45\textwidth]{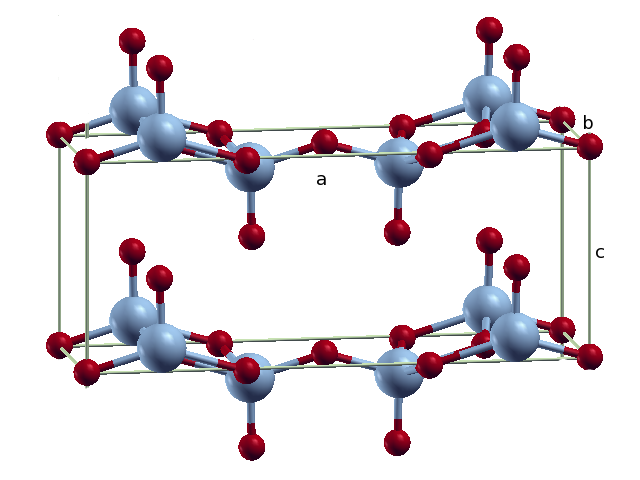}
\caption{The structure of V$_2$O$_5$ bulk ($\alpha$ phase). The box shows the primitive unit cell
that contains two formula units. The 14 atoms of the unit cell are shown, as well as a number of
the atoms of the neighboring unit cells. V atoms are represented by large spheres and O atoms by
small spheres. The labels $a$, $b$, and $c$ indicate the vectors that
span the orthorhombic unit cell. The cleavage plane is perpendicular to the $c$ direction.
Figure created using \textsc{xcrysden}.\protect\cite{xcrysden}}
\label{fig:bulk}
\end{center}
\end{figure}

The ($\alpha$-)V$_2$O$_5$ bulk has orthorhombic symmetry and a layered structure: each vanadium atom is connected 
to five oxygen atoms to create pyramids that share their corners in building a double chain. 
The chains are connected along the edges to form layers that are then stacked to form the
bulk structure  (Figure \ref{fig:bulk}). 

There are three structurally different oxygen atoms in each layer. One is coordinated to one vanadium atom, 
the second is found in a bridging position and the third has a threefold coordinated position. 
The binding of the atoms inside a layer is strong whereas the interactions that keep the layers 
stacked are weaker, resulting in V$_2$O$_5$ being easily cleaved.  

%%%%%%%%%%%%%%%%%%%%%%%%%%%%%%%%%%%%%%%%%%%%%%%%%%%%%%%%%%%%%%%%%%%%%%%%%%%%%%%%%%%%%%%%%%%%%%
\section{Computational methods}

For determining the bulk structure of V$_2$O$_5$ we use the first-principles nonlocal
functional vdW-DF within DFT.\cite{Dion}
We calculate the vdW-DF energies in a post-GGA procedure.\cite{nsc_enough}

The main interest here is on determining the 
 optimal inter-layer spacing in V$_2$O$_5$
(which is the same as the unit cell side length $c$, Figure \ref{fig:bulk}) and
the binding energy, defined as the energy gained  by stacking layers of V$_2$O$_5$. 
We calculate the vdW-DF total
energy $E^\vdW$ for a number of values of $c$ and find the minimum of the energy curve.
For each point on the energy curve the
procedure requires first a sc-GGA calculation, from
which the GGA-based total energy $E^\GGA_\tot$ and the sc-GGA electron density $n$
are evaluated.
Then $n$ is used for evaluating the long-range correlation
contribution arising from the vdW interactions, $E_c^\nl$.
Following the systematic procedure described in more detail in several other 
publications,\cite{PAHgraphite,PEgg,NTgg}
and summarized below, we combine the sc-GGA and nonlocal results to obtain $E^\vdW$.

This section contains three parts explaining the computational methods used.
The first part deals with the sc-GGA calculations used as a basis for obtaining $n$ and some terms of the 
energy  $E^\vdW$.
The second part contains a short summary of the scheme used for the calculation of $E_c^\nl$ and
the vdW-DF total energy $E^\vdW$, and
the third part briefly explains the effect and importance of choice of exchange functional for use in vdW-DF.

\subsection{The sc-GGA calculations}

The sc-GGA calculations are carried out using the plane-wave code\cite{dacapo} \textsc{dacapo} 
with ultrasoft pseudopotentials (USPP).
We describe V$_2$O$_5$ by an orthorhombic unit cell containing two formula units, periodically
repeated in all directions.
The energy cutoff for the expansion of the wave functions is set to 500 eV.
The sc-GGA calculations are carried out using the PBE (Ref.\ \onlinecite{PBE}) flavor of the
GGA for the exchange and correlation functional.
The fast Fourier transform (FFT) grid is chosen such as to have a distance less than 0.12~{\AA} between
nearest-neighbor grid points. The choice of this relatively dense electron density grid is
important for the quality of the subsequent evaluation of the nonlocal correlation contribution.
The Brillouin zone of the unit cell is sampled according to the Monkhorst-Pack scheme 
by means of a 2$\times$4$\times$4 $k$-point sampling.

Our results are converged with respect to the wave function cut off, the number of 
$k$ points needed, and a number of code-internal parameters. 
By lowering the number of $k$ points in the $c$ direction we notice an insignificant 
($\sim 0.001$ eV) change in the minimum value of the sc-GGA total energy curve.

Since the intraplanar bonds have an ionic and partly covalent nature\cite{Kempf} we 
assume the layers to remain rigid when changing the interlayer spacing $c$.
This means that the atoms are kept fixed in their positions (found at the PBE binding distance)
relative to the layer as the distance between the layers is changed, 
with fixed atom-atom distances within the layers.
By analysing the residual Hellmann-Feynman forces on the atoms,
calculated from the GGA-based electron density, we find that this is a reasonable 
approximation.

In this paper we define the binding energy $E_b$ of V$_2$O$_5$ to be the energy 
gained by moving together isolated layers of V$_2$O$_5$ to form the bulk structure.
Each unit cell in the bulk contains two formula units in one layer, periodically repeated, 
and the reference (``layers far apart") 
calculation is carried out with vacuum added on both sides of the layer.
The reference calculation is for a unit cell that has the $c$ direction side length  
four times that of the original unit cell, periodically repeated in all 
directions.\cite{Ziambaras} We use the same reference unit cell for the nonlocal correlation 
($E_c^\nl$) reference calculations and this imposes additional constraints on the 
construction of the reference unit cell, as described further below.

To supplement our analysis of the electron density with a better description of the
core electrons we carry out a set of additional 
GGA calculations using the all-electron DFT code \textsc{gpaw}\cite{gpaw} that is based on 
projector augmented waves\cite{Blochl} (PAW). In these calculations we use settings of
computational parameters as close as possible to the parameters we use in \textsc{dacapo}.  

\subsection{vdW density functional}

We use the vdW-DF scheme to include the vdW interactions in a systematic manner. The
correlation energy $E_c$ is split\cite{IJQC} into a nearly-local part $E_c^0$ and a part that
includes the most nonlocal interactions $E_c^\nl$,
\begin{equation}\label{eq:1}
E_c = E_c^0+E_c^\nl \,.
\end{equation}
The splitting of the correlation contributions makes it possible to employ different approximations
for each term.
In a homogeneous system the term $E^0_c$ is the correlation $E^\LDA_c$ obtained from the local
density approximation (LDA),
and in general\cite{Dion} we approximate $E^0_c$ by $E^\LDA_c$.
The $E_c^\nl$ vanishes for a homogeneous system.
It describes the coupling through the electrodynamic field, the dispersion interaction.
The difference in $E_c^\nl$ contributions provides a description of the interaction which
acts across large distances. The difference is not much influenced by local variations in the
electron density.  Rather, it is
susceptible to the more coarse-grained response of the environment.
The form of $E_c^\nl$ is derived in Ref.~\onlinecite{Dion} and is
\begin{equation}\label{eq:2}
E_c^\nl[n] =
\frac{1}{2}\int\int d\mathbf{r}\,d\mathbf{r}'\,n(\mathbf{r})\phi(\mathbf{r},\mathbf{r}')n(\mathbf{r}')
\end{equation}
where $\phi$ is a kernel, explicitly stated in Ref.~\onlinecite{Dion}.

The electron density $n$ that enters (\ref{eq:2}) is taken from the sc-GGA calculations 
and is described on a grid (the FFT grid of the sc-GGA calculation). 
Our code for evaluating $E_c^\nl$ is not periodic. However, as described in 
detail in Refs.\ \onlinecite{PAHgg,Ziambaras,NTgg}, the natural periodicity within the bulk unit cell is easily 
represented by explicitly including a number of the periodic images of the unit cell, with the nearby 
grid points carefully described and the grid points far away only described via a coarse version of the grid. 

Similar to the energy contributions from the sc-GGA calculations the term $E_c^\nl$ must be
evaluated relative to a reference calculation with the V$_2$O$_5$ layers ``far apart''. 
As described above, this is here achieved with a unit cell four times the original unit cell in the $c$ 
direction, with the V$_2$O$_5$ layer placed close to the middle of the unit cell. For this reference
calculation of $E_c^\nl$ no periodicity in the $c$ direction is assumed. The layer must be placed such   
that locally, with respect to the (FFT) grid points
on which $n$ is described, the nuclei maintain the same positions in the bulk and the reference calculation.
This means that the (FFT) grid must be chosen such that the bulk and the reference calculations have the exact same
grid point separation; in this study we need precisely four times the number of FFT points in the reference
calculation compared to the original bulk calculation. 

The combination of correlation terms in (\ref{eq:1}) avoids double counting of correlation contributions.
Using the correlation energy $E_c$ from (\ref{eq:1}) the total energy in the vdW-DF scheme is
\begin{equation}\label{eq:3}
E^\vdW=E^\GGA_\tot-E_c^\GGA+E_c^\LDA+E_c^\nl,
\end{equation}
where E$^\GGA_\tot$ is the total energy from the sc-GGA calculation and $E_c^\GGA$
and $E_c^\LDA$ the GGA respective LDA correlation calculated from the sc-GGA electron density~$n$. 
The layer binding
energy $E_b$ (per unit cell) we define as the energy gained by moving the layers in V$_2$O$_5$ together 
accordion-like,\cite{Ziambaras} 
\begin{equation}
E_b=-(E^\vdW_{\mbox{\scriptsize bulk}}-E^\vdW_\reff).
\end{equation}

\subsection{Exchange functionals}
With every choice of approximation for correlation functional comes a need for choosing a suitable
flavor of exchange functional.
In previous work we have chosen to combine the correlation $E_c$ of (\ref{eq:1}) with
the exchange from the GGA flavor revPBE.\cite{revPBE}
This choice has been the default for some time because calculations\cite{Dion,langrethjpcm2009} 
have shown that revPBE$_x$\cite{xforx} does not
give rise to spurious binding. This choice of revPBE$_x$ is thus conservative in that it ensures that all the 
improvement in binding compared to GGA calculations comes from 
the improvement in correlation (\ref{eq:1}).

However, it has also been shown that revPBE$_x$ is overly repulsive in the binding region of many
vdW-bonded systems.\cite{HFcomment} 
In the present paper our aim is to study the binding of the layers of V$_2$O$_5$ by 
qualitatively examining the improvement in binding in the vdW-DF calculations over the usual GGA
calculations. 
We do not here aim for a ``perfect" description of the binding in the sense of best 
possible numerical fit to experiment. Rather, by showing the results of using a few choices of sound 
exchange flavors, we illustrate the sensitivity to those choices and find the range of values of $c$ and $E_b$
within which we expect the physical values to be. 
A forthcoming publication\cite{group} addresses and discusses choices of exchange 
flavors with vdW-DF in extended (layered) systems.
In the present presentation of V$_2$O$_5$ binding we pick two promising candidates for
exchange to use with vdW-DF, besides the well known PBE$_x$ and
revPBE$_x$ functionals.
These are C09$_x$ (Ref.\ \onlinecite{Cooper}) and PW86$_x$ (Ref.\ \onlinecite{aboutPW86}).

\begin{figure}
\begin{center}
\includegraphics[width=0.47\textwidth]{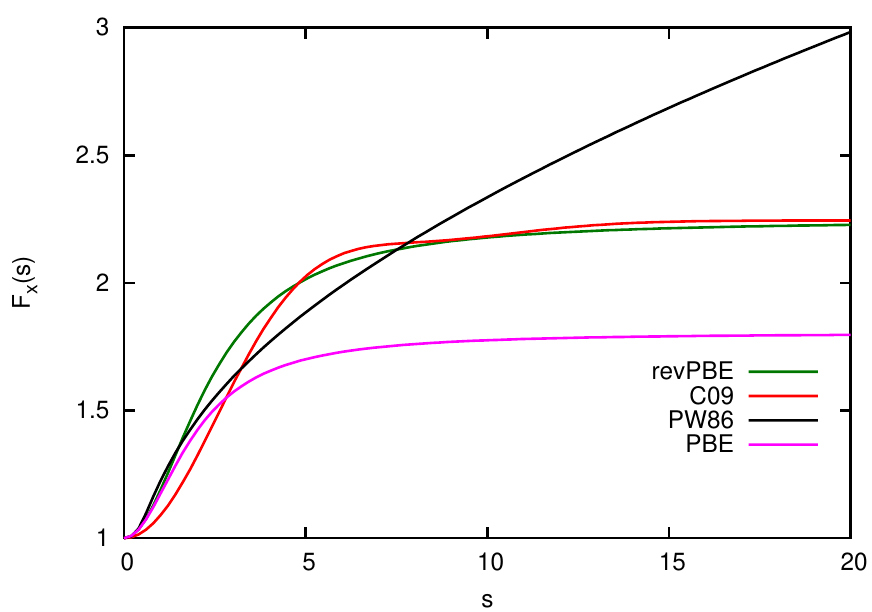}
\caption{\label{fig:fx}(Color) Exchange functional enhancement factors $F_x$ as functions of the reduced 
density gradient $s \sim |\nabla n|/n^{4/3}$. Shown are the exchange flavors used here.} 
\end{center}
\end{figure}

GGA exchange may be described by the enhancement over LDA exchange in terms of the  
enhancement factor $F_x$ 
\begin{equation}
E_x = \int d\mathbf{r} \, n(\mathbf{r})\, \epsilon_x^\LDA\left(n(\mathbf{r})\right)\, F_x\left(s(\mathbf{r})\right)
\label{eq:fxint}
\end{equation} 
where $\epsilon_x^\LDA=-c_x n^{1/3}$ with $c_x=3^{4/3}/(4\pi^{1/3})$
is the LDA exchange energy density 
and $s=c_s |\nabla n|/n^{4/3}$ with $c_s=1/(2\pi^{2/3} 3^{1/3})$ is the reduced electron density gradient. 
The $F_x(s)$ curves for the
exchange flavors considered here are shown in Fig.\ \ref{fig:fx} for a range of $s$-values.

Dense materials have significant electron densities in most regions of the material and relevant 
values of $s$ for those systems lie mainly in the range\cite{perdew2005} $0<s<3$. 
For sparse materials, with regions of very low electron 
densities, values of $s>3$ also play a role and cannot be ignored. As is clear from Fig.\ \ref{fig:fx}
the various flavors of exchange give different contributions at large values of $s$ and the choice of 
exchange flavor is therefore more important for sparse than for dense matter. 

In practice we include a particular flavor of exchange functional into the vdW-DF by 
subtracting the PBE$_x$ energy contribution $E_x^\PBE$ from the total energy $E_\tot^\GGA$
of the sc-GGA calculations
(because the sc-GGA calculations are carried out with the 
PBE flavor of GGA) and adding the value of the relevant exchange energy, 
calculated from the sc-GGA electron density $n$.

%%%%%%%%%%%%%%%%%%%%%%%%%%%%%%%%%%%%%%%%%%%%%%%%%%%%%%%%%%%%%%%%%%%%%%%%%%%%%%%%%%%%%%%%%%%%%% 
\section{Results and discussion}

A number of previous DFT studies of the V$_2$O$_5$ bulk system have used GGA calculations for 
obtaining the lattice constants of the orthorhombic unit cell.%
\cite{Braithwaite,Pirovano2007,Pirovano,willinger2004,GPMV,Goclon,Xiao,Reeswinkel} The in-plane lattice 
parameters, in the present paper called $a$ and $b$ (but in parts of the literature called 
$a$ and $c$), are found to be close to the experimental values. Table \ref{tab:GGAlatt} lists
for comparison the results of two experiments as well as the GGA results of Ref.\ \onlinecite{Pirovano}. 
The inter-plane binding distance is the same as the third lattice parameter (here $c$,
in parts of the literature $b$). For $c$ a number of groups have found significantly larger values  
than the experimental value. \cite{Pirovano2007,Pirovano,willinger2004,Goclon,Xiao,Kresse01}
Ganduglia-Pirovano and Sauer\cite{Pirovano} carried out GGA calculations using PW91\cite{PW91} with 
PAW, emphasising convergence and accuracy. This resulted
in a $c$ lattice constant even further away from the experimental value than the values from 
previous GGA calculations. 

The deficiency in the description of the binding from GGA calculations 
has been attributed to the presence of vdW interactions between the V$_2$O$_5$ layers.\cite{Pirovano}
It is well known that GGA does not adequately describe vdW interactions,\cite{IJQC} thus if 
the layers of V$_2$O$_5$ are mostly bound to each other by vdW forces it is not surprising that GGA calculations
do not give physically reasonable results. In the following we report our results from GGA calculations, 
from vdW-DF calculations, and discuss the effect of exchange and the binding character of V$_2$O$_5$.  

\subsection{GGA results}
Exploring the V$_2$O$_5$ bulk system first with the (\textsc{dacapo}-based) sc-GGA calculations
we find, similar to the studies mentioned above, values of the $a$ and $b$ lattice
constants in good agreement with experiment and a value for $c$ about 11\% too large when
compared to experiment (Table \ref{tab:GGAlatt}). 
The PBE total energy curve is shallow in the $c$ direction. In order to
determine the minimum with any reasonable accuracy we first find the approximate 
minimum, and then evaluate the total energy at 125 points in ($a$, $b$, $c$) parameter 
space in a small region around the approximate minimum. A three-dimensional 
polynomial fit to those points yields an energy minimum for the values of $a$, $b$, and $c$ 
as listed in Table \ref{tab:GGAlatt} along with the results of our supplemental \textsc{gpaw}
calculations.

The shallow minimum of the PBE total energy curve at $c=4.87$ {\AA} has the value 0.18 eV 
per unit cell compared to the layers taken apart, $c\rightarrow\infty$ 
($c=4.89$~{\AA} with 0.21 eV per unit cell for our \textsc{gpaw} calculations).
This is in good agreement with the PW91 results of 
Ref.\ \onlinecite{Pirovano} where $c=4.84$ {\AA} and a similar small surface energy in the cleavage plane
was found. For truncated bulk they found the surface energy 0.048 J/m$^2$, approximately corresponding to a 
binding energy per unit cell of 0.25 eV. Our GGA results are also in good agreement with the 
PAW-based PBE results of Ref.\ \onlinecite{Pirovano2007}.
In all four GGA calculations, the value of the binding energy is unphysically small,
illustrating the deficiency of GGA.

\begin{table}
\caption{\label{tab:GGAlatt} Equilibrium lattice parameters ($a$, $b$, and $c$) and 
interlayer binding energy per unit cell ($E_b$). 
Numbers in parenthesis are obtained by PW91
calculations and used in the calculations indicated. 
For the vdW-DF calculations the flavor of the exchange
functional used is shown explicitly, with subscript $x$ denoting the exchange part of the
exchange-correlation functional.} 

\begin{tabular}{l ccccc}
  \hline\hline
                                &$a$         &$b$      &$c$     & $E_b$  & $C_{33}$ \\
                                &[\AA]       &[\AA]    & [\AA]  & [eV]   & [GPa]\\
      \hline
      \textit{This work} \\
      PBE with USPP             & 11.52     & 3.57    & 4.87    & 0.18   & 24 \\
      PBE with PAW$^{\mathrm{a}}$& (11.55)  &(3.58)   & 4.89    & 0.21   & 21 \\
      vdW-DF, revPBE$_x^{\mathrm{b}}$&(11.55)&(3.58)  & 4.72    & 0.84   & 23 \\
      vdW-DF, revPBE$_x$        &(11.55)    &(3.58)   & 4.72    & 0.86   & 22 \\
      vdW-DF, PBE$_x$           &(11.55)    &(3.58)   & 4.46    & 1.28   & 57 \\
      vdW-DF, PW86$_x$          &(11.55)    &(3.58)   & 4.46    & 1.19   & 61 \\
      vdW-DF, C09$_x$           &(11.55)    &(3.58)   & 4.28    & 1.10   & 58 \\
\hline
      \textit{Comparison} \\
      PW91 with PAW$^{\mathrm{c}}$& 11.55   & 3.58    & 4.84    & 0.25   & - \\
      Experiment$^{\mathrm{d}}$ & 11.512    & 3.564   & 4.368   & -      & - \\
      Experiment$^{\mathrm{e}}$ & 11.508    & 3.559   & 4.367   & -      & - \\
      \hline\hline
\multicolumn{6}{l}{$^{\mathrm{a}}$Using the DFT code \textsc{gpaw};}\\
\multicolumn{6}{l}{$^{\mathrm{b}}$revPBE$_x$ extracted from the \textsc{dacapo} code;}\\
\multicolumn{6}{l}{$^{\mathrm{c}}$Ref.\ \protect\onlinecite{Pirovano}, $E_b$ is here estimated as the cleavage energy,}\\
\multicolumn{6}{l}{calculated from the given surface energy of truncated}\\
\multicolumn{6}{l}{surfaces: 0.048 J/m$^2$;}\\ 
\multicolumn{6}{l}{$^{\mathrm{d}}$Ref.\ \protect\onlinecite{Enjalbert};} \\
\multicolumn{6}{l}{$^{\mathrm{e}}$Ref.\ \protect\onlinecite{Chou}, Powder of V$_2$O$_5$ ($\sim 420$ nm crystals).}
    \end{tabular}
\end{table}

\subsection{vdW-DF results}

\begin{figure}[bth]
\begin{center}
\includegraphics[width=0.47\textwidth]{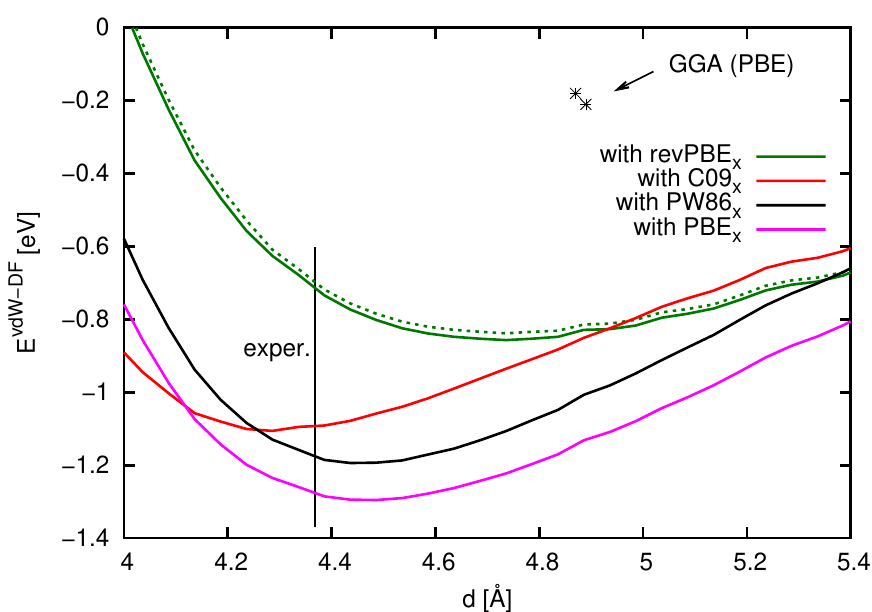}
\caption{\label{fig:Ecurves} $E^\vdW$ using various flavors of exchange.
For revPBE$_x$ we plot both the result of using the exchange energy
provided directly by \textsc{dacapo} (dashed line) and using our external
implementation (solid lines). 
The two stars indicate the optimal points of our
two PBE-based GGA calculations and the vertical line the experimentally found value of $c$.}
\end{center}
\end{figure}

With vdW-DF we find total-energy curves (at fixed values of $a$ and $b$)
with enhanced binding compared to GGA (Fig.~\ref{fig:Ecurves}).
The position of the minimum of the total-energy curve depends on our
choice of GGA exchange flavor, as discussed in the previous section.
The optimal values of $c$ and the binding energies for each choice of
exchange functional are listed in Table \ref{tab:GGAlatt}.

For vdW-DF with C09$_x$ the value of $c$ is close to the experimental value,
in fact the optimal value of $c$ with C09 is 2\% \textit{smaller\/} than the
experimental value. PBE$_x$, and PW86$_x$ as recommended in Ref.\ \onlinecite{murray2009}, 
improve the binding over the results of GGA by decreasing $c$ about 
$0.5$ {\AA}, resulting in a value of $c$ that deviates from experiment by only  
2\%. The binding energies range from 0.8 eV/unit cell
(for revPBE$_x$) to 1.3 eV/unit cell (PW86$_x$). As expected,  
revPBE$_x$ leads to longer binding distances (9\% larger than experiment) 
and a smaller binding energy than use of the other exchange flavors
because revPBE$_x$ is overly repulsive. 

In Table \ref{tab:GGAlatt} we list two results based on revPBE$_x$. 
Previously, we have used the values of the revPBE$_x$ 
energy that were provided from sc-GGA \textsc{dacapo} calculations for  
input to the vdW-DF calculations. In this work we rely on our
own external implementation for the exchange energy calculations.  
The implementation differences between  \textsc{dacapo} and our external code
are minor, and as shown in Fig.\ \ref{fig:Ecurves} 
for revPBE$_x$, the differences in the results are also minor, and can be ignored.  
 
While the discussion of best and physically most reasonable
choice of exchange flavor to go with vdW-DF is still open,  
it is clear from our total-energy calculations that the vdW
forces do indeed play a prominent role in the binding of the V$_2$O$_5$ layers.
Earlier, another group\cite{sauer2008} has examined the role of the vdW-forces in 
V$_2$O$_5$ by means of the semi-empirical DFT-D approach (Ref.\ \onlinecite{grimme}),
arriving at a similar conclusion. 

The vdW-DF total energy curves in Fig.\ \ref{fig:Ecurves} have small wiggles, in particular in 
the expansion region ($c > 4.6$ {\AA}). We find that the wiggles mainly arise from the exchange part of the 
energy, as also seen in, e.g., Refs.\ \onlinecite{benzene,PAHgg}.
The minimum of the $E^\vdW$ curves is found by a polynomial fit to the points closest to the minimum.
{}From this fit we also extract the corresponding values of the elastic coefficient $C_{33}$ (Table  \ref{tab:GGAlatt}).
Because V$_2$O$_5$ is more soft (easier to expand or compress) in the direction perpendicular to
the layers, compared to other directions, the value of $C_{33}$ is also a good estimate of the
bulk modulus of the system. 
Based on the likely most realistic exchange choices (i.e., excluding here the overly repulsive revPBE$_x$)
we find that V$_2$O$_5$ bulk with  $C_{33}\approx 60$ GPa is a 
somewhat more stiff material than graphite, which has\cite{graphiteC33} $C_{33}= 37$--$41$ GPa.

\subsection{Numerical noise and exchange functionals}
Our DFT calculations are based on a description of the valence electron wavefunctions by USPP. 
The USPP are not normconserving 
and tend to give ``noise" in regions of very low
electron density, yielding some small (unphysical) negative values of $n$.
In order to handle these small negative values of $n$, \textsc{dacapo} replaces
on the FFT grid points all values $n<n_{\floor,\dac}$ by $n_{\floor,\dac}$ before calculating the
exchange energy. The floor is given in terms of the Bohr radius  $a_0$,
 $n_{\floor,\dac}=10^{-10}$ $|e|/a_0^3\approx 10^{-9}$ $|e|/${\AA}$^3$ being 
a small but positive minimum value of the density.

Points in space in which this floor is applied become, via the formula $s\sim |\nabla n|/n^{4/3}$, 
points with extremely large (unphysical) values of $s$. For exchange choices like PBE$_x$ and revPBE$_x$ with a constant 
asymptote of $F_x(s)$ the effect of this replacement is small:  
Very large values of $s$ ($\gg 100$) contribute much less to the energy integral (\ref{eq:fxint})
than do moderate values of $s$ ($\approx 5-10$), because $F_x(s)$ is basically constant at large $s$
(small $n$) and the integrand $n^{4/3}F_x$ thus vanishes for small $n$. 
However, for more ``aggressive" exchange functionals,
in terms of growth of $F_x$ for large values of $s$, this may cause problems in material systems 
with large regions of small 
or vanishing electron density (``vacuum"), such as our reference calculations.\cite{codecomment}
This is further discussed in a forthcoming publication.\cite{group} Here, we simply state that the modification
used in our external implementation is to instead \textit{remove\/} contributions to the 
integral of (\ref{eq:fxint}) that come from 
points in space with $n< n_\floor=10^{-15}$ $|e|/${\AA}$^3$, and that the procedure is not very sensitive
to the particular value of $n_\floor$. 

At the time when dense, bulk-like systems were the main systems examined with DFT, relevant values of 
$s$ were considered to be small, in the approximate range\cite{PBE} $0 < s < 3$.
In sparse matter this is not always the case: in our sparse matter systems we routinely 
work with (physical) values of $s$ which are several orders of magnitude larger than this. 
In our V$_2$O$_5$ reference calculations, consisting of a single layer of  V$_2$O$_5$
surrounded by vacuum, almost 50\% of the spatial grid points give rise to $s$-values 
larger than 12 both in our  USPP-based electron density from the \textsc{dacapo}
calculations and in the pseudo part of the electron density of our supplemental
\textsc{gpaw} calculations. In comparison, the bulk calculations at close-to-experimental 
separation ($c=4.3$ {\AA}) show mostly small values of $s$.
We find that $s=2.7$ is the largest value in  the \textsc{gpaw} calculation 
for bulk V$_2$O$_5$ while only 0.4\% of 
spatial points have large values ($s>12$) in the corresponding \textsc{dacapo} calculation. 

As a measure of the effect of removing all points with $n < n_\floor$ (including points with negative 
values of $n$) we calculate an error measure $\eta$. This error measure is 
defined as the (absolute value of the) contributions that would have come 
from the excluded points, weighted with the unit cell size  $V=abc$,  
\begin{equation}
\eta=\frac{1}{V}\, \Delta\!\!\! \int_{n<n_\mathrm{floor}} \!\!\!\!\!\!
d\mathbf{r}\, |n| \, \epsilon_x^\LDA\left(|n|\right)\, F_x\left(c_s \frac{|\nabla n|}{|n|^{4/3}}\right)
\end{equation}
where the difference indicated by $\Delta$ is between contributions from bulk and reference calculations. 
We find that this measure is approximately constant over the range of layer separations studied here, with values
ranging from 0.5 meV/{\AA}$^3$ for C09$_x$ to 1.0  meV/{\AA}$^3$ for PW86$_x$. The fact that $\eta$ is approximately 
constant with $c$ fits well with our expectation that spacial points with $n < n_\floor$ appear in the vacuum region of the 
unit cell in the reference calculations, a region that expands approximately as fast as $V$.   

We note that the numerical problems faced here are similar to those one of us has  
previously documented for a molecular dimer.\cite{benzene} 
Then, as part of an early vdW-DF study, we investigated
the energy difference between a far-apart benzene dimer in
a unit cell of a certain (large) size and two isolated benzene molecules each in a unit cell of 
the same size as the original unit cell. Whereas physically this energy difference should 
vanish, a small but finite contribution $\delta E_\reff$ appeared.\cite{benzene} This unphysical difference
$\delta E_\reff$ would not converge with, e.g., planewave 
energy cut off or other convergence parameters and it was found to originate mainly from 
the exchange part of the energy.   
In dimer and adsorption studies this problem can be  
overcome by simple error cancellation using a fixed size of the unit cell and moving the 
material fragments around.\cite{benzene,PAHgg} In bulk calculations the unit cell size must necessarily change, and
no such full error cancellation is possible, leading to a usually small but nonvanishing $\eta$. 

\begin{figure*}
\begin{center}
\includegraphics[height=7 cm]{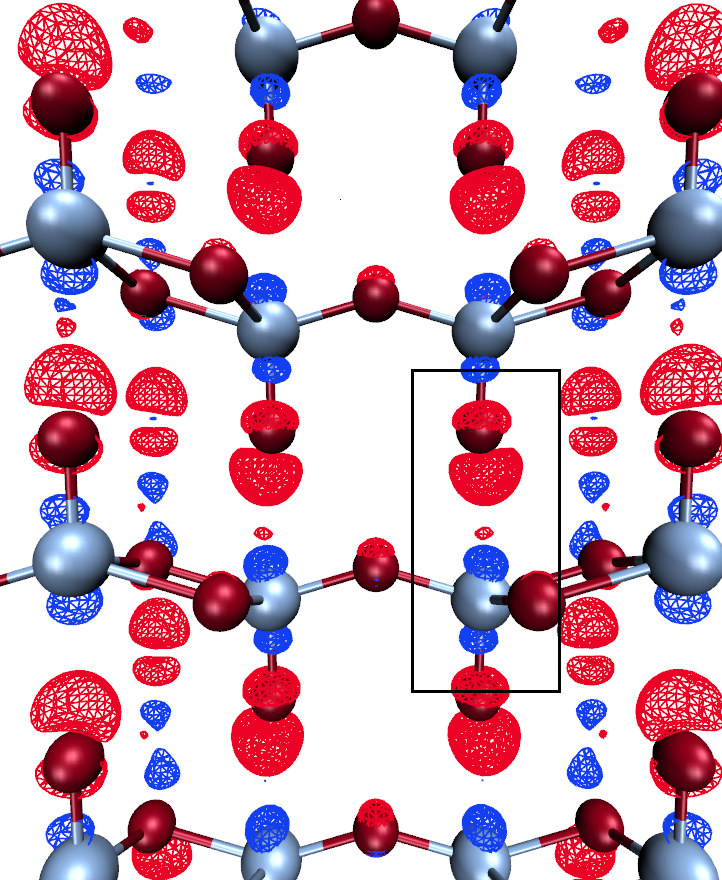} \hspace{1.5 cm} \includegraphics[height=7 cm]{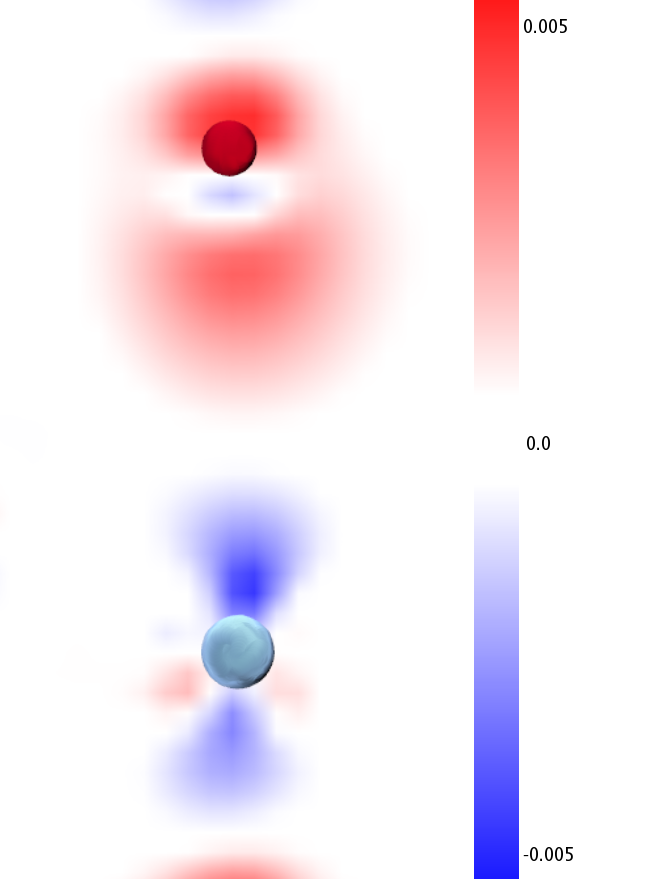}
\caption{(Color) Change in electron density, $\Delta n(\mathbf{r})$, as a result of assembling the  layers of V$_{2}$O$_{5}$. 
Left:  Isosurfaces of values 0.02 $|e|/${\AA}$^3$ are shown, red indicates \textit{gain\/} of electrons (loss of 
charge). The box indicates the region of the plot shown in the right panel. 
Right: The electron 
density difference is shown in a slice through the system, through the positions of the V atom and the
corresponding vanadyl O atom of the next layer.
Color scales are indicated in the right-hand bar, in units of $|e|/${\AA}$^3$. Figure created with 
\textsc{vmd}.\protect\cite{vmd}}
\label{fig:chdchangeslice}
\end{center}
\end{figure*}

\subsection{Binding of layers} 
The results of our vdW-DF calculations, regardless of choice of exchange functional, indicate that 
the vdW forces are indeed important in the V$_2$O$_5$ system. 
We have further analysed the system by extracting the change in electron density, $\Delta n(\mathbf{r})$,
that arises when moving an isolated layer of  V$_2$O$_5$ into the bulk structure (atom positions
kept fixed). This short qualitative analysis is based on the $n(\mathbf{r})$ extracted from the sc-GGA calculations,
but we expect no major deviations from the result of using an electron density based on sc vdW-DF
calculations. 

Fig.\ \ref{fig:chdchangeslice}
shows that only a small amount of electron charge is moved when the V$_2$O$_5$ layers bind.
Most of this change is on the vanadyl O atom and the V atom.  
A Bader analysis,\cite{Bader1990} using the algorithm 
described in Ref.\ \onlinecite{Henkelman} and discussed in Ref.\ \onlinecite{Borck2006},
confirms this picture of very little change of charge: only about 0.1 electrons move from V to vanadyl O,
the charge on all other atoms is almost unchanged. This is in agreement with results of 
Ref.\ \onlinecite{Laubach}. 

We agree that the static electron density $n(\mathbf{r})$ may not be the best tool to interpret 
the binding character, as suggested and discussed in Ref.\ \onlinecite{Sture}. Nevertheless, we believe that
qualitatively the combined interpretation from Figure \ref{fig:chdchangeslice} and from the Bader analysis is sufficient
to conclude that no short-range binding type (like ionic or covalent bonds) can explain the main part of the
binding of the  V$_2$O$_5$ layers. 
We find that inclusion of dispersive interaction is essential to complete the description of  V$_2$O$_5$
cohesion.

%%%%%%%%%%%%%%%%%%%%%%%%%%%%%%%%%%%%%%%%%%%%%%%%%%%%%%%%%%%%%%%%%%%%%%%%%%%%%%%%%%%%%%%%%%%%%%
\section{Summary}
In this paper we have studied the role of the van der Waals bonding inside a layered 
oxide by means of the (by now) well tested vdW-DF method.\cite{Dion,Thonhauser} In particular we have underlined 
the importance of accounting for the long range interactions that characterize this 
bonding in order to recover a structure of V$_2$O$_5$ close to that found by experiments. 
We have shown that the 
flavor of the exchange functional chosen for the vdW-DF calculations can give noticeable 
differences in calculated binding distances and energies and the definition of the most 
appropriate one is still a matter of debate. Specifically, in this paper we have tested 
two promising functionals to be used with vdW-DF: C09$_x$ and PW86$_x$. 
We have also analysed effects of numerical noise that may arise in studying sparse matter 
within DFT.
This is important because there are large regions with low density, regions that may cause 
errors in the evaluation of exchange. Finally, we have identified how this error varies 
with the choice of exchange flavor.

\acknowledgments
We thank M.V.\ Ganduglia-Pirovano for introducing us to the problem of 
describing V$_2$O$_5$ with DFT and for constructive discussions. 
We also thank J.\ Rohrer, K.\ Berland, and P.\ Hyldgaard for 
valuable discussions of exchange functional issues.
Partial support from the Swedish Research Council (VR) is gratefully acknowledged,
as well as allocation of computer time at UNICC/C3SE (Chalmers) and
SNIC (Swedish National Infrastructure for Computing). 

%%%%%%%%%%%%%%%%%%%%%%%%%%%%%%%%%%%%%%%%%%%%%%%%%%%%%%%%%%%%%%%%%%%%%%%%%%%%%%%%%%%%%%%%%%%%%%


\begin{thebibliography}{99}

\bibitem{Keller}
B.M. Weckhuysen and D.E. Keller,
Catal. Today \textbf{78}, 25  
(2003) and references therein.

\bibitem{Julien}
C. Julien, E. Haro-Poniatowski, M.A. Camacho-L\'opez, L. Escobar-Alarc\'on, and 
J. J\'{\i}menez-Jarqu\'{\i}n,
Mater. Sci. Engin. B \textbf{65}, 170 (1999).

\bibitem{Sabana} M.B. Sahana, C. Sudakar, C. Thapa, G. Lawes, V.M. Naik, 
R.J. Baird, G.W. Auner, R. Naik, and K.R. Padmanabhan,
Mater. Sci. Engin. B \textbf{143}, 42 (2007).

\bibitem{Chou} S.-L. Chou, J.-Z. Wang, J.-Z. Sun, D. Wexler, M. Forsyth, H.-K. Liu,
D.R. MacFarlande, and S.-X. Dou,
Chem. Mater. \textbf{20}, 7044 (2008).

\bibitem{Braithwaite}
J.S. Braithwaite, C.R.A. Catlow, J.D. Gale, and J.H. Harding,
Chem. Mater. \textbf{11}, 1990 (1999).

\bibitem{GPMV}
M.V. Ganduglia-Pirovano and J. Sauer,
J. Phys. Chem. B \textbf{109}, 374 (2005).

\bibitem{Reeswinkel}
T. Reeswinkel, D. Music, and J.M. Schneider,
J. Phys.: Condens. Matter \textbf{21}, 145404 (2009).

\bibitem{Goclon}
J. Goclon, R. Grybos, M. Witko, and J. Hafner,
J. Phys.: Condens. Matter \textbf{21}, 095008 (2009).

\bibitem{Xiao}
Z.R. Xiao and G.Y. Guo,
J. Chem. Phys. \textbf{130}, 214704 (2009).

\bibitem{willinger2004} 
M. Willinger, N. Pinna, D.S. Su, and R. Schl\"ogl, Phys. Rev. B
\textbf{69}, 155114 (2004).

\bibitem{Pirovano2007}
J.L.F. Da Silva, M.V. Ganduglia-Pirovano, and J. Sauer, Phys. Rev. B
\textbf{76}, 125117 (2007).

\bibitem{Pirovano}
M.V. Ganduglia-Pirovano and J. Sauer,
Phys. Rev. B \textbf{70}, 045422 (2004).

\bibitem{Kresse01}
G. Kresse, S. Surnev, M.G. Ramsey, and F.P. Netzer,
Surf. Sci. \textbf{492}, 329 (2001).

\bibitem{Li}
Z.-Y. Li and Q.-H. Wu,
J. Mater. Sci.: Mater. Electron. \textbf{19}, S366 (2008).

\bibitem{Goclon2}
J. Goclon, R. Grybos, M. Witko, and J. Hafner,
Phys. Rev. B \textbf{79}, 075439 (2009).

\bibitem{Hejduk}
P. Hejduk, M. Witko, and K. Hermann,
Top. Catal. \textbf{52}, 1105 (2009).

\bibitem{layerPRL} 
H. Rydberg, M. Dion, N. Jacobson, E. Schr\"oder, P. Hyldgaard, 
S.I. Simak, D.C. Langreth, and B.I. Lundqvist,
Phys. Rev. Lett. \textbf{91}, 126402 (2003). 

\bibitem{grimme}
S. Grimme, Journ. Comput. Chem. \textbf{27}, 1787 (2006).

\bibitem{sauer2008}
T. Kerber, M. Sierka, and J. Sauer,
Journ. Comput. Chem. \textbf{29}, 2088 (2008).

\bibitem{Dion}
M. Dion, H. Rydberg, E. Schr\"oder, D.C. Langreth, and B.I. Lundqvist,
Phys. Rev. Lett. \textbf{92}, 246401 (2004); \textbf{95}, 109902(E) (2005).

\bibitem{Thonhauser}
T. Thonhauser, V.R. Cooper, S. Li, A. Puzder, P. Hyldgaard, and D.C. Langreth,
Phys. Rev. B \textbf{76}, 125112 (2007).

\bibitem{NTgg} J. Kleis, E. Schr\"oder, and P. Hyldgaard,
Phys. Rev. B \textbf{77}, 205422 (2008).

\bibitem{berland2010}
K. Berland and P. Hyldgaard,
J. Chem. Phys. \textbf{132}, 134705 (2010).

\bibitem{xcrysden}
A. Kokalj, Comp. Mater. Sci. \textbf{28}, 155 (2003).

\bibitem{nsc_enough}
This is reasonable since the differences
in the resulting atomic positions and total energies to those of a
fully self-consistent vdW-DF calculation
have been shown to be negligible (Ref. \protect\onlinecite{Thonhauser}).

\bibitem{PAHgraphite}
S.D. Chakarova-K\"ack, E. Schr\"oder, B.I. Lundqvist, and
D.C. Langreth, Phys. Rev. Lett. \textbf{96}, 146107 (2006).

\bibitem{PEgg}
J. Kleis, B.I. Lundqvist, D.C. Langreth, and E. Schr\"oder
Phys. Rev. B \textbf{76}, 100201(R) (2007).

\bibitem{dacapo}
Open-source plane-wave DFT computer code \textsc{dacapo},
\texttt{http://wiki.fysik.dtu.dk/dacapo/}.

\bibitem{PBE}
J.P. Perdew, K. Burke, and M. Ernzerhof,
Phys. Rev. Lett. \textbf{77}, 3865 (1996).

\bibitem{Kempf}
Y.L. Kempf, B. Silvi, A. Dietrich, C.R.A. Catlow, and B. Maigret,
Chem. Mater. \textbf{5}, 641 (1993).

\bibitem{Ziambaras}
E. Ziambaras, J. Kleis, E. Schr\"oder, and P. Hyldgaard,
Phys. Rev. B \textbf{76}, 155425 (2007).

\bibitem{gpaw}
Open-source, grid-based PAW-method DFT code \textsc{gpaw},
\texttt{http://wiki.fysik.dtu.dk/gpaw/}.

\bibitem{Blochl} P.E. Bl\"ochl, Phys. Rev. B \textbf{50}, 17953 (1994).

\bibitem{IJQC}
D.C. Langreth, M. Dion, H. Rydberg, E. Schr\"oder, P. Hyldgaard, and B.I. Lundqvist,
Intern. J. of Quantum Chem. \textbf{101}, 599 (2005).

\bibitem{PAHgg} S.D. Chakarova-K\"ack, A. Vojvodic, J. Kleis, P. Hyldgaard, and E. Schr\"oder,
New Journ. Phys. \textbf{12},  013017 (2010).

\bibitem{revPBE}
Y. Zhang and W. Yang,
Phys. Rev. Lett. \textbf{80}, 890 (1998).

\bibitem{langrethjpcm2009}
D.C. Langreth, B.I. Lundqvist, S.D. Chakarova-K\"ack, V.R. Cooper, M. Dion, P. Hyldgaard, 
A. Kelkkanen, J. Kleis, L. Kong, S. Li, P.G. Moses, E. Murray, A. Puzder, H. Rydberg, 
E. Schr\"oder, and T. Thonhauser,
Journ. Phys.: Cond. Matter \textbf{21}, 084203 (2009).  

\bibitem{xforx}
By subscript $x$ we denote the exchange part of the density functionals.

\bibitem{HFcomment}
Even though Hartree-Fock exchange in principle
provides exact exchange, it is not a perfect alternative either,
as discussed in for example Refs. \protect\onlinecite{IJQC,perdew2005,murray2009}.

\bibitem{group} E. Londero et al., unpublished. 

\bibitem{Cooper}
V.R. Cooper, Phys. Rev. B \textbf{81}, 161104(R) (2010).

\bibitem{aboutPW86} As defined in Ref.\ \protect\onlinecite{PW86} with the refitted values given
in Ref.\ \protect\onlinecite{murray2009}.

\bibitem{perdew2005}
J.P. Perdew, A. Ruzsinszky, J. Tao, V.N. Staroverov, G.E. Scuseria, and G.I. Csonka,
Jour. Chem. Phys. \textbf{123}, 062201 (2005).

\bibitem{PW91}
J.P. Perdew, in \textit{Electronic Structure of Solids '91,} edited by P. Ziesche 
and H. Eschrig (Akademie Verlag, Berlin, 1991), p. 11;
J.P. Perdew, J.A. Chevary, S.H. Vosko, K.A. Jackson, M.R. Pederson, D.J. Singh, 
and C. Fiolhais, Phys. Rev. B \textbf{46}, 6671 (1992); Phys. Rev. B \textbf{48}, 4978(E) (1993).

\bibitem{Enjalbert}
R. Enjalbert and J. Galy,
Acta Cryst. C \textbf{42}, 1467 
(1986).

\bibitem{murray2009} {\'E}.D. Murray, K. Lee, and D.C. Langreth,
Jour. Chem. Theor. Comput. \textbf{5}, 2754 (2009).

\bibitem{benzene} S.D. Chakarova and E. Schr\"oder,
Mater. Sci. Engin. C \textbf{25}, 787 (2005).

\bibitem{graphiteC33} \textit{Landolt-B\"ornstein Search} (Springer-Verlag, Berlin,
2003), http://link.springer.de.

\bibitem{codecomment}We note that while we here show numerical problems
encountered using USPP with
the DFT code \textsc{dacapo} some other DFT codes may
face similar problems at various levels. Most DFT codes were not written with sparse systems
and small energies in mind and we recommend an explicit test of effects of numerical noise
with vdW-DF calculations.

\bibitem{vmd} W. Humphrey, A. Dalke, and K. Schulten,
Journ. Molec. Graphics \textbf{14}, 33 (1996).

\bibitem{Bader1990} R.F.W. Bader, Atoms in Molecules---A Quantum Theory
(Oxford University Press, Oxford, 1990).

\bibitem{Henkelman}
G. Henkelman, A. Arnaldsson, and H. J\'onsson,
Comput. Mater. Sci. \textbf{36}, 254 (2006).

\bibitem{Borck2006} {\O}. Borck and E. Schr\"oder,
Journ. Phys.: Condens. Matter \textbf{18}, 10751 (2006).

\bibitem{Laubach}
S. Laubach, P.C. Schmidt, A. Thi{\ss}en, F.J. Fernandez-Madrigal, Q.-H. Wu,
W. Jaegermann, M. Klemm, and S. Horn,
Phys. Chem. Chem. Phys. \textbf{9}, 2564 (2007).

\bibitem{Sture}
S. Nordholm and W. Eek,
Intern. J. Quantum Chem. \textbf{110}, xx (2010) [DOI: 10.1002/qua.22490].

\bibitem{PW86}J.P. Perdew and Y. Wang, Phys. Rev. B \textbf{33}, 8800(R) (1986).
%not Wang and Perdew as written in jctc09.
 
\end{thebibliography}
\end{document}